\documentclass[fleqn,usenatbib,usedcolumn]{mnras}

\usepackage{mathptmx}

\usepackage[T1]{fontenc}
\usepackage{ae,aecompl}

\usepackage{graphicx}	
\usepackage{amsmath}	
\usepackage{amssymb}	
\usepackage[export]{adjustbox}

\newcommand{\myemail}{lindsay.fuller@utsa.edu}
\newcommand{\degree}{$^{\circ}$}
\newcommand{\um}{$\mu$m}

\title[31.5 \um~imaging observations of AGN]{Investigating the dusty torus of Seyfert galaxies using SOFIA/FORCAST photometry}

\author[L. Fuller et al.]{L. Fuller$^{1}$\thanks{E-mail: \myemail}, E. Lopez-Rodriguez$^{2,3}$, C. Packham$^{1,4}$, C. Ramos-Almeida$^{5,6}$\thanks{Ram\'on y Cajal Fellow}
\newauthor A. Alonso-Herrero$^{1,7,8}$, N.~A.~Levenson$^{9}$, J. Radomski$^{10}$, K. Ichikawa$^{4}$ 
\newauthor I. Garc\'ia-Bernete$^{5,6}$,O. Gonz\'alez-Mart\'in$^{11}$, T. D\'iaz Santos$^{12}$, M. Mart\'inez-Paredes$^{11}$ 
\\
\\
$^{1}$Department of Physics \& Astronomy, University of Texas at San Antonio, One UTSA Circle, San Antonio, TX 78249, USA \\
$^{2}$Department of Astronomy, University of Texas at Austin, 2515 Speedway, Stop C1402, Austin, Texas 78712, USA \\
$^{3}$McDonald Observatory, University of Texas at Austin, 2515 Speedway, Stop C1402, Austin, Texas 78712-1206, USA \\
$^{4}$National Astronomical Observatory of Japan, 2-21-1 Osawa, Mitaka, Tokyo 181-8588, Japan \\
$^{5}$Instituto de Astrof\'isica de Canarias, Calle V\'ia L\'actea s/n, E-38205, Tenerife, Spain \\
$^{6}$Universidad de La Laguna, Departamento de Astrof\'isica, E-38206 La Laguna, Tenerife, Spain \\
$^{7}$Centro de Astrobilogia (CAB,CSIC-INTA), ESAC Campus, E-28692 Villanueva de la Ca\~nada, Madrid, Spain \\
$^{8}$Department of Physics, University of Oxford, Oxford OX1 3RH, UK	\\
$^{9}$Gemini Observatory, Casilla 603, La Serena, Chile \\
$^{10}$SOFIA/USRA, NASA Ames Research Center, Moffett Field, CA 94035, USA \\
$^{11}$Instituto de Radioastronomía y Astrofísica (IRyA-UNAM), 3-72 (Xangari), 8701, Morelia, Mexico \\
$^{12}$Núcleo de Astronomía de la Facultad de Ingeniería, Universidad Diego Portales, Av. Ejército Libertador 441, Santiago, Chile}

\date{Accepted XXX. Received YYY; in original form ZZZ}

\pubyear{2016}

\begin{document}
\label{firstpage}
\pagerange{\pageref{firstpage}--\pageref{lastpage}}
\maketitle

\begin{abstract}

We present 31.5 \um~imaging photometry of 11 nearby Seyfert galaxies observed from the Stratospheric Observatory For Infrared Astronomy (SOFIA) using the Faint Object infraRed CAmera for the SOFIA Telescope (FORCAST).  We tentatively detect extended 31 \um~emission for the first time in our sample.  In combination with this new data set, subarcsecond resolution $1-18$ \um~imaging and $7.5-13$ \um~spectroscopic observations were used to compute the nuclear spectral energy distribution (SED) of each galaxy.  We found that the turnover of the torus emission does not occur at wavelengths  $\leq$31.5 \um, which we interpret as a lower-limit for the wavelength of peak emission.  We used \textsc{Clumpy} torus models to fit the nuclear infrared (IR) SED and infer trends in the physical parameters of the AGN torus for the galaxies in the sample.  Including the 31.5 \um~nuclear flux in the SED 1) reduces the number of clumpy torus models compatible with the data, and 2) modifies the model output for the outer radial extent of the torus for 10 of the 11 objects.  Specifically, six (60\%) objects show a decrease in radial extent while four (40\%) show an increase.  We find torus outer radii ranging from $<$1pc to 8.4 pc.
\end{abstract}




\section{Introduction}

The unified model \citep{A1993,UP1995} of active galactic nuclei (AGN) posits that all AGN are essentially the same type of object viewed from different lines of sight.  This orientation-based model depends on a circumnuclear toroidal region of optically and geometrically thick dust which can obscure a central region of high-energy emission (a super massive black hole of $\sim$10$^{6-9} M_{\odot}$ and accretion disk), responsible for producing high energy photons.  Observed broad (FWHM $\sim$ 10$^{3}$ - 10$^{4}$ km/s) and narrow (FWHM $<$ 10$^{3}$ km/s) emission line features in Type 1 AGN are due to a direct view of the central engine, whereas the circumnuclear dust torus obscures a direct view of the central engine and broad line emission region in Type 2 AGN.  Strong support for the unification scheme came from spectropolarimetric observations of NGC 1068 \citep{AM1985}, a Type 2 Seyfert galaxy.  It was shown that NGC 1068 contains polarized broad emission lines that are hidden from direct view, but clearly revealed in polarized radiation, proving that this Seyfert 2 has properties similar to a Seyfert 1.  Subsequently, evidence for a hidden broad line region was also found in several other highly polarized Seyfert 2's \citep[e.g.][]{MillGood,Brindle1990,Tran}.  In both Seyfert 1 and 2 galaxies, dust grains in the torus absorb optical and ultraviolet radiation from the central engine and re-radiate in the infrared (IR).  

It was assumed early on that most of the dust in the torus must have been distributed in molecular dust clouds or would otherwise not survive expected temperatures \citep{KB1988}; hence early models assumed a homogeneous distribution of dust \citep{PK1992,GD1994,EfRR1995,Granato1997,Sieb2004} for its computational simplicity.  In the case of homogeneous models, the amount of mid-IR (MIR) to far-IR (FIR) emission suggested a torus outer extent of up to a few hundred parsecs.  These models were based on moderate spatial resolution ($\gg$1 arcsec) photometry, and thus suffered from contamination from diffuse dust emission and stellar emission in the core of the host galaxy.  High spatial resolution MIR imaging observations on 8-m class telescopes quickly ruled out a torus of such large extent.  Specifically, N- and Q-band observations from Gemini South  provided an upper limit on the outer radius of $\sim$2 pc for Circinus \citep{Packham2005} and 1.6 pc for Centaurus A \citep{Radomski2008}.  Further N-band interferometric observations on VLTI/MIDI \citep{Jaffe2004,Tristram2007,Raban2009,B2013} and more recently sub-millimeter observations from ALMA \citep{GB2016} confirmed a smaller radius for several objects.    

The small size of the torus is effectively modeled by an inhomogeneous, ``clumpy" dust distribution throughout its volume.  In this scenario, dust of differing temperatures can exist at the same radius \citep{Nenkova2002,Schart2005}, i.e. the illuminated face of one cloud (emitting in near-IR (NIR)) can exist at the same radius as the shadowed face of another cloud (emitting in MIR).  Models assuming a clumpy distribution \citep{Nenkova2002,Nenkova2008,Nenkova2008b,Honig2006,Stalevski2012,Sieb2015} also account for the variety of spectral energy distributions \citep[SEDs;][]{Fadda,AH2003,RA09,RA11,AH2011,Lira2013,Ichi2015} that are seen.  For example, the NIR to MIR SED is sensitive to the inclination angle of the torus and its width, with the hotter dust within the torus contributing to flatten the SED.  The outer radius is best constrained by FIR emission \citep{RA11b,AR2013} since temperatures are generally cooler further away from the central engine, though in detail depends on the total dust distribution.  Consequently, knowing the wavelength of peak emission from the torus gives insight as to its radial size.  Some models show peak IR emission from the torus between $\sim$$20 - 40$ \um~\citep{Nenkova2008,H2010,Mull2011,Feltre2012}.  Thus, observations at wavelengths longer than 20 \um~are essential in determining the wavelength of peak emission.

Observing AGN at the highest possible spatial resolution ensures minimal contamination of the signal from the surrounding diffuse dust and stellar emission in the core of the host galaxy.  To optimally constrain the torus model parameters, several authors \citep{RA09,H2010,RA11,AH2011,Ichi2015} have combined subarcsecond-resolution $1 - 20$ \um~observations to evaluate the SED.  Although these previous studies have effectively described the parameters taking into account high spatial resolution observations at wavelengths $\lambda < 25$ \um, the lack of high spatial resolution observations at wavelengths $\lambda > 25$ \um~leaves the SED at longer wavelengths largely unconstrained, reducing the quality of the fitting to the clumpy models.  Unfortunately, since the atmosphere is opaque in FIR, large aperture observations from the ground are impossible at these wavelengths.  The space-based \textit{Spitzer} telescope covers this range, but the diameter (0.85m) severely limits its resolution ($\sim$9.1 arcsec at 30 \um).  NASA's Stratospheric Observatory For Infrared Astronomy (SOFIA) presents a unique solution for non-space-based observations in the FIR.  The 2.5-m telescope on board the aircraft is three times the size of \textit{Spitzer}, providing a much better spatial resolution ($\sim$3.4 arcsec at 30 \um).

In this paper, we present new 31.5 \um~imaging data from NASA's SOFIA telescope for a sample of 11 nearby Seyfert galaxies.  We used the \textsc{Clumpy} torus models of \citet{Nenkova2008,Nenkova2008b} and a Bayesian approach \citep{AR2009} to fit the IR ($1-31.5$ \um) nuclear SEDs in order to constrain torus model parameters.  We aim to determine the AGN contribution to the flux within the SOFIA aperture, estimate the potential turnover of the IR torus emission, and determine the effect on model fits after adding the extended wavelength range. Section \ref{OBS} of this paper contains the observations and data reduction; Section \ref{results} contains our photometric analysis method and its results; Section \ref{MOD} gives the results from the model fitting with the addition of our 31.5 \um~photometric point; Section \ref{DIS} contains an analysis of the torus model parameters; and Section \ref{CON} gives our conclusions.



\vspace{-0.7cm}

\section{Observations and Data Reduction} 
\label{OBS}

\subsection{Sample Selection} 

The sample of this pilot study was selected based on two basic criteria: 1) well-known, bright, nearby Seyfert galaxies previously studied \citep{RA09, RA11, AH2011, Ichi2015} and well-sampled in the $1-18$ \um~ regime with sub-arcsecond spatial resolution, and 2) bolometric luminosities in the range 42 $\leq$ $L_{bol}$ $\leq$ 45 erg $s^{-1}$.  As a result, 11 sources were selected; their properties are given in Table \ref{redshift}.

\begin{table}
\centering

\caption{AGN sample}
\label{redshift}
{\footnotesize
\begin{tabular}{p{1.4cm}p{0.6cm}p{0.5cm}cp{0.8cm}p{1.1cm}p{0.5cm}}
\hline

Object		&	Type			&	$z$		&	Distance	&	Scale 		&	log L$_{bol}$	&	Ref(s)\\
			&					&				&	(Mpc)		&	\scriptsize{(pc/ arcsec)}	&	(erg s$^{-1}$)	&			\\		

\hline
MCG-5-23-16	& Sy2		& 0.0085	& 34.5		&	167 		&	44.4		&	a,1		\\
Mrk 573		& Sy2		& 0.0170	& 69.2		&	336		&	44.4		&	b,2		\\
NGC 2110		& Sy2		& 0.0076	& 31.0		&	150		&	43.9		&	c,2		\\
NGC 2992	& Sy1.9		& 0.0077	& 31.3		&	152 		&	43.5		&	d,3		\\
NGC 3081	& Sy2		& 0.0080	& 32.4		&	157 		&	43.6		&	e,2		\\
NGC 3227	& Sy1.5		& 0.0039	& 15.7		&	76   		&	43.1		&	f,4		\\
NGC 3281	& Sy2		& 0.0107	& 43.2		&      212 		&	43.8		&	e,2		\\
NGC 4388	& Sy2		& 0.0047	& 19.0		&	92 		&	44.1		&	g,2		\\
NGC 5506	& Sy1.9		& 0.0062	& 25.1		&	122 		&	44.3		&	h,2		\\
NGC 7469	& Sy1		& 0.0163	& 66.3		& 	322		&	44.6		&	b,4		\\
NGC 7674	& Sy2		& 0.0289	& 117.5		&	570 		&	45.5		&	b,2		\\
\hline
\end{tabular}\\
\textsc{References.} AGN Type: a)  \citet{BWW1990}, b) \citet{OM1993}, c) \citet{R2003}, d) \citet{G2000}, e) \citet{PCB1983}, f) \citet{DD1988}, g) \citet{Kuo2011} h) \citet{N2002}. Redshift: \citet{deV1991}.  Distances were calculated from the redshift using $H_{0}$=73.8 km s$^{-1}$Mpc$^{-1}$.  Luminosity: 1) \citet{AH2011}, 2) \citet{Marinucci2012}, 3) \citet{GB2015}, 4) \citet{RA11}.
}
\end{table}

\subsection{Photometry}

Observations (Proposal ID: \#002\_35, PI: Lopez-Rodriguez) were made using the Faint Object Infrared Camera for the SOFIA Telescope (FORCAST; \citealp{Hertel2012}) on the 2.5-m SOFIA telescope \citep{Young2012}. FORCAST is a dual-channel IR camera and spectrograph sensitive in the wavelength range of $5-40$ \um. Each channel consists of 256 $\times$ 256 pixels with a pixel scale of 0.768 arcsec pixel$^{-1}$, providing an effective field of view (FOV) of 3.4 $\times$ 3.2 arcmin, after correcting for distortion.  The two channels - the short wavelength camera (SWC), $\lambda < 25$ \um, and the long-wavelength camera (LWC), $\lambda > 25$ \um~- can be used simultaneously or individually.
 
The 31.5 $\mu$m filter ($\Delta\lambda=$ 5.7\um) was used with the LWC in single channel mode.  The 31.5 \um~filter provides the best angular resolution and sensitivity for the suite of FORCAST filters available at wavelengths longer than 30 \um. Observations were made using the two-position chop-nod (C2N) method with symmetric nod-match-chop (NMC) to remove time-variable sky background and telescope thermal emission, and to reduce the effect of 1/$f$ noise from the array. The chop throw was 1 arcmin in all observations. A summary of the observations is shown in Table \ref{Summary}. 

Data were reduced using the \textsc{forcast\_redux} pipeline v1.0.1beta following the method described by \citet{Hertel2012} to correct for bad pixels, ``droop'' effect, non-linearity, and cross-talk. We found that for our data the merging stage produces high background variations affecting the photometric measurements; thus, the merging stage was not included in the data reduction.  

\begin{table*}
\centering
\caption{Summary of SOFIA observations}
\label{Summary}
\begin{tabular}{cccccccc}
\hline
Object & Observation date & Altitude &On-source & Chop rate & Chop angle & Mission ID &Calibration stars \\
   	& (yyyymmdd)	&  (ft) 	& 	Time (s) 	&	(Hz) & (deg)	& 	&	 \\
	
\hline
MCG-5-23-16	& 20140503	& 38000	& 324	& 5.09	 & 180	& FO\_F167 	& $\beta$ UMi, $\alpha$ Boo, $\sigma$ Lib \\
Mrk 573		& 20150205	& 43000	& 384	& 4.31	& 150	& FO\_F192	& $\alpha$ Boo						  \\
NGC 2110		& 20150205	& 43000	& 786	& 4.31	& 150	& FO\_F192	& $\alpha$ Boo						  \\
NGC 2992	& 20140502	& 39000	& 232	& 5.09       & 60		& FO\_F166	& $\beta$ UMi, $\alpha$ Boo, $\sigma$ Lib \\
NGC 3081	& 20140502	& 39000	& 480	& 4.71 	 & 150	& FO\_F166	& $\beta$ UMi, $\alpha$ Boo, $\sigma$ Lib \\
NGC 3227	& 20140506	& 38000	& 152	& 4.62	 & 120	& FO\_F166	& $\beta$ UMi, $\alpha$ Boo, $\sigma$ Lib \\
NGC 3281	& 20140502	& 39000	& 300	& 5.41	 & 150	& FO\_F166	& $\beta$ UMi, $\alpha$ Boo, $\sigma$ Lib \\
NGC 4388	& 20140502	& 38000	& 162	& 4.62	 & 150	& FO\_F166	& $\beta$ UMi, $\alpha$ Boo, $\sigma$ Lib \\
NGC 5506	& 20140502	& 38000	& 261	& 5.09	 & 150	& FO\_F166	& $\beta$ UMi, $\alpha$ Boo, $\sigma$ Lib \\
NGC 7469	& 20140604	& 43000	& 231	& 3.97 	 &  60	& FO\_F176	& $\alpha$ Boo, $\beta$ And \\
NGC 7674	& 20140604 	& 43000	& 165	& 3.97	 & 150	& FO\_F176	& $\alpha$ Boo, $\beta$ And \\
\hline

\end{tabular}		
\end{table*}

The point spread function (PSF) of the observations was estimated as the co-average of the set of standard stars associated with the observing run (Table \ref{Summary}). The FWHM of the co-averaged standard star was estimated to be 3.40$\pm$0.12 arcsec, in excellent agreement with the best measurement of the FWHM of 3.4  arcsec quoted  in the SOFIA Observer's Handbook v3.0.0\footnote{The FWHM of the filter suite for FORCAST can be found at: \url{http://www.sofia.usra.edu/Science/ObserversHandbook}.}. The galaxy images were flux-calibrated using the set of standard stars of the observing run, following the method described by \citet{Hertel2012}.  The calibration factor is given in units of a photo-electron rate, Me$^{-}$ s$^{-1}$ Jy$^{-1}$.  The final calibration factor was estimated as the average of the individual calibration factors for each standard star, and corrected by the zenith angle of the observations for each galaxy. Two galaxies could have compromised flux measurements due to: (1) high array bias, $>1.3$, that could affect the flux measurements of NGC 7674 or (2) low zenith angle, $<32$\degree, observations that could produce vignetting in NGC 4388.  

\subsection{Compilation of NIR/MIR Photometry and MIR Spectroscopy for SED}

We compiled the highest angular resolution NIR (Table \ref{NIRphot}) and MIR photometry (Table \ref{MIRphot}) and MIR spectroscopy (Table \ref{Spectroscopy}) from the literature to construct the nuclear $1.2 - 31.5$ \um~SEDs of each AGN in our sample.  The SEDs have already been successfully used by previous authors \citep{RA09,H2010,RA11,AH2011,Ichi2015} in giving physical information about the torus using models.  The compiled NIR photometry includes observations from IRCam3/UKIRT, NICMOS/HST, NACO/VLT, and IRAC1/ESO with angular resolution ranging from 0.09 arcsec to 0.7 arcsec with L band photometry of NGC 2110 as an exception (we use an upper limit).  All of the MIR photometry was obtained using large ground-based telescopes (i.e. Gemini North/South and VLT) with angular resolution in the range 0.3 arcsec to 0.6 arcsec.  MIR spectroscopy was also obtained using 8-m class telescopes (Gemini North/South, VLT, GTC) with similar resolution.  We add the 31.5 \um~photometric data to these SEDs to study the effect of the new measurement on the physical parameters of the torus.

\begin{table*}
\centering
\begin{minipage}{280mm}
\caption{High spatial resolution nuclear NIR fluxes.}
\label{NIRphot}

\begin{tabular}{ccccccc}
\hline

Object			&	\multicolumn{5}{c}{NIR Flux Densities (mJy)}		&	Ref(s).	 		\\
				&	J band		&	H band		&	K band		&	L band		&	M band		& 		\\	
\hline
MCG-5-23-16		& $1.1\pm0.3$		& $3.7\pm0.9$  	& $10.7\pm2.7$	& $79.5\pm16.0$ 	& $139.4\pm28.0$ 	& 	 a    	\\

Mrk 573 			& $0.15\pm0.06$	& $0.54\pm0.04$	& $3.2\pm0.6$		& $18.8\pm3.8$	&  $41.3\pm8.3$	& 	 b,c,a\\

NGC 2110			& \dots 			&	\dots 		&     \dots			&	<33			& \dots			&	a	\\	
								
NGC 2992		& \dots 			&  $1\pm0.1$  	 	& $2.8\pm0.6$	 	& $22.7\pm4.5$ 	& $35.7\pm7.1$   	& 	a       \\
				
NGC 3081		& \dots 			& $0.22\pm0.04$	& \dots 			&\dots  			&\dots  			& 	c        \\

NGC 3227		& \dots 			& $7.8\pm0.8$	 	& $16.6\pm1.7$	& $46.7\pm9.3$	& $72\pm27$		&  d,b,e      \\
			
NGC 3281		& \dots 			& $1.3\pm0.2$	 	& $7.7\pm0.8$	 	& $103\pm9$		& $207\pm25$		&    f  	 \\

NGC 4388		& $0.06\pm0.02$	& $0.71\pm0.28$	& \dots  			& $39.9\pm8.0$	& \dots			&  b            \\

NGC 5506		& $13\pm3$ 		& $53\pm8$		& $80\pm12$	 	& $290\pm44$		& $530\pm106$	&  g,a   	  \\

NGC 7469		& $8.0\pm1.6$		& $15\pm2.3$		& $20\pm3$		& $84\pm13$		& $96\pm14$	        &  g 	    	  \\

NGC 7674		& $1.0\pm0.25$	& $5.0\pm0.5$		& $12.3\pm3.1$	& $53\pm11$		& $108.6\pm10.8$	& a     	 \\

\hline
\end{tabular}\\
\textsc{References:} a) \citet{AH2001}, b) \citet{AH2003}, c) \citet{Q2001}, \\
d) \citet{K2007}, e) \citet{W1987}, f) \citet{S1998}, g) \citet{P2010}.
\end{minipage}
\end{table*}

\begin{table*}
\centering
\begin{minipage}{280mm}
\caption{High spatial resolution nuclear MIR fluxes.}
\label{MIRphot}
\begin{tabular}{ccccccccc}
\hline
Object		&	\multicolumn{4}{c}{MIR Flux Densities (mJy)}												&	Ref(s)			\\
			&	N band 								&	Filter					&	Q band 	&  Filter	&	 				\\
\hline
MCG-5-23-16	&	$358.2\pm16.7$;  $633.4\pm24.0$			&	VISIR/ArIII,PAH2\_2	&	$1540\pm7$	&	VISIR/Q2		&  a,b  	\\
Mrk 573		&	$177\pm27$							& 	T-ReCS/N				&	$415\pm104$	&	T-ReCS/Qa		&	c	\\
NGC 2110		&	$169.1\pm6.4$;   $286\pm28$;	 $294.3\pm9$	& 	VISIR/PAH1,PAH2\_2;Michelle/N' &	$519\pm28$	&	VISIR/Q1		&	a,d,e\\
NGC 2992	&	$175\pm26$  							& 	Michelle/N'	&	$521\pm130$				&	Michelle/Qa	&	c	\\
NGC 3081	&	$83\pm12$ 							& 	T-ReCS/Si-2	&	$231\pm58$				&	T-ReCS/Qa	&	c	\\
NGC 3227	&	$180\pm11$;  $320\pm22$;  $401\pm60$		&   	VISIR/ArIII,PAH2\_2;Michelle/N'     &	\dots	 		&	\dots			&	a,c	\\
NGC 3281	&	$355\pm8$ 							&  	T-ReCS/N		&	$1110\pm278$				&	T-ReCS/Qa	&	c 	\\
NGC 4388	&	$195\pm29$  							& 	Michelle/N'	&	$803\pm201$				&	Michelle/Qa	&	c	\\
NGC 5506	&	$873\pm131$ 							&	Michelle/N'	&	$2200\pm550$				&	Michelle/Qa	&	c	\\
NGC 7469	&	$174\pm26$;  $530\pm80$ 				&	T-ReCS/Si-2;VISIR/PAH2\_2	&	$1354\pm339$	&	T-ReCS/Qa	&	f,g	\\
NGC 7674	&	$518\pm22$	 						&	VISIR/NeII			&	 \dots			&	\dots			&	a	\\

\hline
\end{tabular}\\
\textsc{References:}, a) \citet{H2010}, b) \citet{R2010}, c) \citet{RA09}, d) \citet{Mason2009},\\
 e) \citet{Asmus2014}, f) \citet{RA11}, g) \citet{P2010}
\end{minipage}
\end{table*}


\begin{table}
\begin{minipage}{70mm}
\caption{High spatial resolution N-band spectroscopy}
\label{Spectroscopy}
\begin{tabular}{lccc}
\hline
Object	&	Instrument	& 	Slit Width	& Ref(s) \\
		&				&	(arcsec)	&		\\
\hline
MCG-5-23-16	&	VISIR		&	0.75		&  a \\
Mrk 573		&	\dots			&			&	\\
NGC 2110		&	Michelle		&	0.36		&  b \\
NGC 2992	&	CanariCam	&	0.52		&  c	\\
NGC 3081	&	T-ReCS		&      0.65		& d	\\
NGC 3227	&	VISIR		&     0.75		&  a \\
NGC 3281	&	T-ReCS		& 	0.35		&  d	\\
NGC 4388	&	CanariCam	& 	0.52		&  c	\\
NGC 5506	&	T-ReCS		&	0.36		&  e  \\
NGC 7469	&	VISIR		&	0.75		&  a	\\
NGC 7674	&	VISIR		&	0.75		&  a	\\
\hline
\end{tabular}
\textsc{References:} a) \citet{H2010}, b) \citet{Mason2009}, 
c) \citet{AH2016}, d) \citet{GM2013}, e) \citet{Roche2007}. \\
\end{minipage}
\end{table}

\section{Imaging Analysis}
\label{results}

\subsection{SOFIA Imaging}

\begin{figure*}
\centering
\includegraphics[scale=0.7]{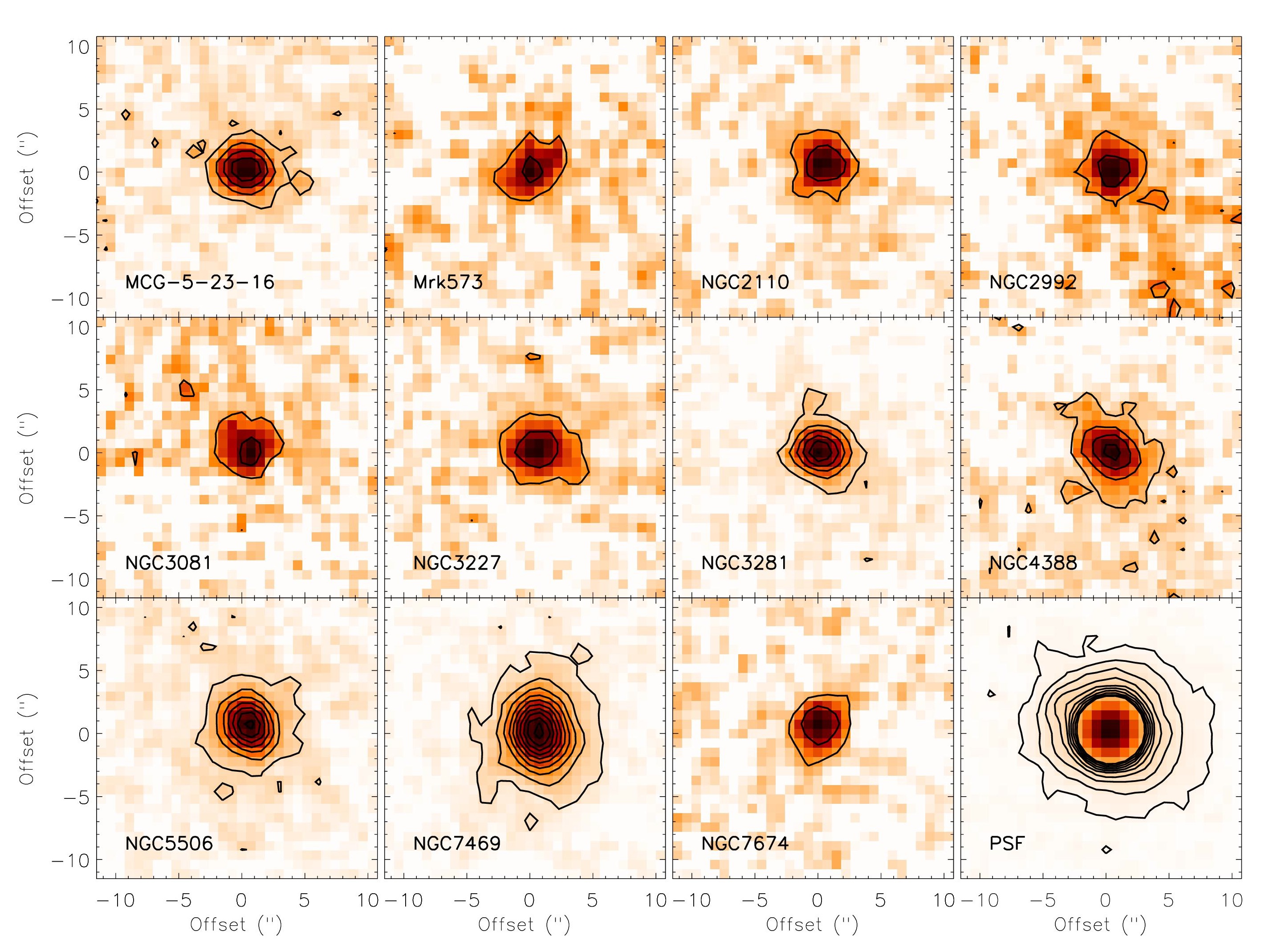}
\caption{SOFIA/FORCAST imaging at 31.5 \um~showing the 20 $\times$ 20 arcsec central region of our AGN sample.  The lowest level contour is 3$\sigma$, then contours step up in intervals of 5$\sigma$.  The PSF follows the same contours; the image indicates the high signal-to-noise ratio.  North is up; east is left.  Offset is measured from the central region of each galaxy in our AGN sample.}
\label{all_agn}
\end{figure*}

Images of the 11 AGN in the 31.5 \um~filter are presented in a 20 $\times$ 20 arcsec field of view (FOV) in Figure \ref{all_agn} along with the PSF of SOFIA; the lowest level contour is 3$\sigma$ and the contours increase in intervals of 5$\sigma$.  These are the first FIR images of these Seyfert galaxies at this resolution.  By comparing the FWHM of the PSF with the FWHM of the images, we found that six of the objects are point sources, while NGC 2992, NGC 4388, NGC 5506, NGC 7469 and possibly NGC 7674 potentially show some extended emission.  The extended emission does not follow the position angle of the chop-throw, though further observations are necessary to confirm these structures.  The study of the extended emission detected by these observations are outside of the scope of this paper; these will be analyzed in a follow up paper with further observations at 37.1 \um~using SOFIA/FORCAST. 

Despite the spatial resolution afforded by SOFIA, diffuse IR emission and star formation can potentially contaminate the nuclear fluxes we aim to obtain.  To minimize contamination of extended emission to the AGN measurement within the 3.4 arcsec FWHM of SOFIA, we used the PSF-scaling method described by \citet{Radomski2003,RA09,Mason2012,GB2015}.  Two different photometric measurements were made for each galaxy.  1) The flux density in a circular aperture of 18 arcsec diameter was measured. The 18 arcsec diameter was used based on the 18 arcsec aperture used for the flux calibration to account for 100\% of the flux of the standard stars.  2) The PSF, representing the maximum flux contribution of the unresolved component, was scaled to the peak of the AGN emission.  Then the flux density was measured in the 18 arcsec aperture around the scaled PSF.

\begin{figure*}
\centering
\includegraphics[scale=0.2]{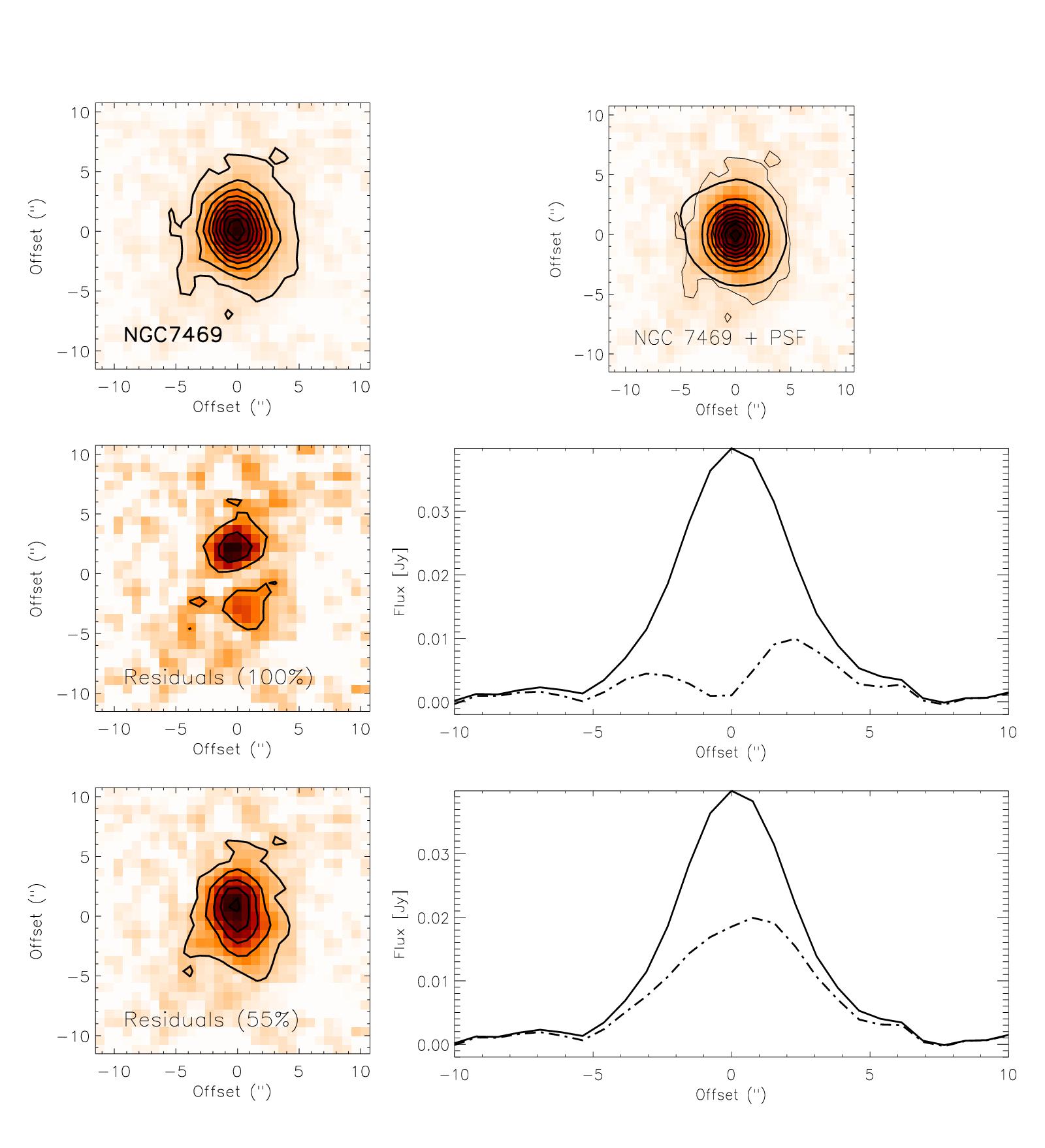}
\caption{Top left: 31.5 \um~image of NGC 7469 as shown in Fig. \ref{all_agn}; Top right: 31.5 \um~image with overlay PSF contours (contours start at the 3$\sigma$ level of the SOFIA image); Middle left: residual image after subtraction of 100\% of the PSF; Middle right: Profiles of the residual and 100\% scaled PSF; Bottom left: residuals after subtraction of 55\% of PSF; Bottom right: Profiles of the residuals and 55\% scaled PSF.  1 arcsecond corresponds to 322 pc.}
\label{psf_fitting}
\end{figure*}

An example of the scaling is shown in Figure \ref{psf_fitting}.  When the PSF is scaled to 100\% of the total emission, then subtracted, the residual image shows some structure.  The residuals, representing emission from the host galaxy, should show a flat profile.  Hence, that description is not physically accurate.  When the PSF is scaled to 55\% of total emission, the profile of the residual flattens, giving a more physically accurate representation of emission from the host galaxy.  

The uncertainty in the photometry was estimated as the variability of the calibration factors of the set of standard stars associated with the observing run, giving an uncertainty in the aperture photometry of 12\% (3$\sigma$). For the PSF-scaling photometry, the 12\% uncertainty due to the variation of the calibration factors, and the 10\% uncertainty induced by a variable PSF obtained by cross-calibrating the standard stars, were added in quadrature.  

\begin{figure*}
\includegraphics[scale=0.7]{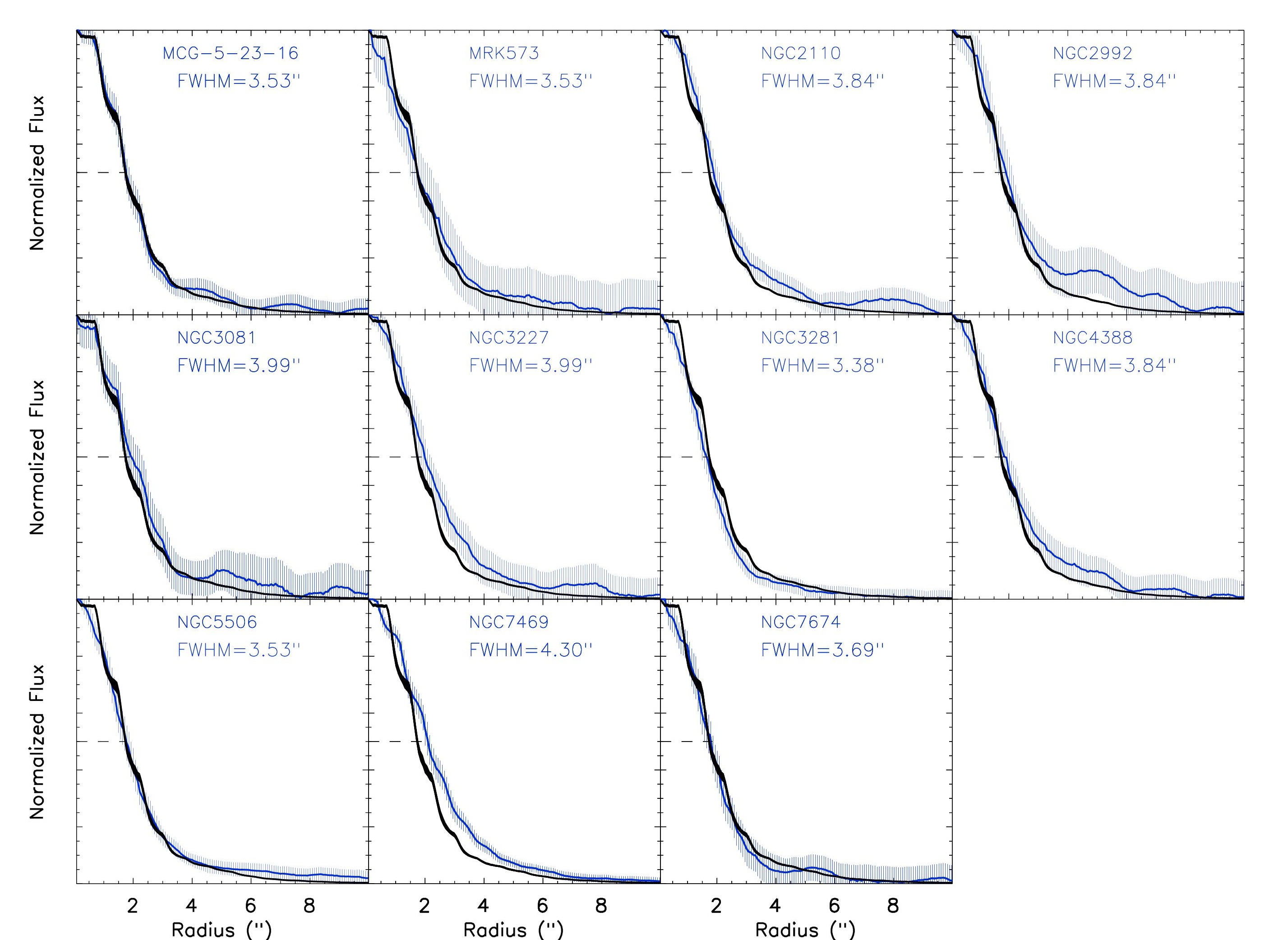}
\caption{Radial profiles of each AGN (solid blue line) are being compared to the averaged PSF of SOFIA (solid black line).  The shaded blue region indicates the uncertainty from the background, as well as from variations in the signal when finding the mean value at increasing radii from the center.  The uncertainty in the PSF measurement is shown as black shading. The dashed horizontal line represents the FWHM determined by direct measurement.  Airy rings are seen in some profiles; maxima occur at 4.3 arcsec and 7.0 arcsec.}
\label{rad_prof}
\end{figure*}

To account for possible flux excess due to high background and/or extended galaxy emission, we analyzed the radial profiles of each AGN within the PSF-scaling aperture.  We first corrected any profiles that showed a non-zero background.  We then estimated the extended flux using measurements within an 18 arcsec diameter, and subtracted any excess from the PSF-scaling photometry.  Figure \ref{rad_prof} shows the radial profile of the PSF (black solid line) as compared to the radial profile of the AGN (blue solid line).  From the radial profiles, extension can be seen most clearly in NGC 2992, NGC 3227, and NGC 7469.  Column 5 of Table \ref{Photometry} shows the contribution of the PSF as a percentage of the total flux in the radial profile image.  

The profiles shown in Figure \ref{rad_prof} show distinct Airy rings for at least two objects - MCG-5-23-16 and NGC 2992.  The maxima of the Airy rings occur at 4.3 arcsec and 7.0 arcsec at the wavelength of our observations.  The combination of errors (shaded blue/black regions) in the AGN and PSF flux measurements are considered to be the standard deviation 1) of the background variation and 2) of the variation in the signal at increasing pixel radii from the center.  Also shown is the FWHM (dashed line) determined by direct measurement.

Table \ref{Photometry} shows the flux of each AGN before accounting for any contamination in the signal, after performing the PSF scaling routine, and also taking into account any excesses from the radial profiles.  First, the total flux from the SOFIA observations within the 18 arcsec aperture (F$_{18"}$) is given in Column 2.  Then the level of PSF scaling (PSF$_{\mbox{\tiny sub}}^{a}$) and subsequent flux measurement (F$_{\mbox{\tiny PSF}}$) are given in Columns 3 and 4, respectively.  Then the PSF flux as a percentage of total flux as evidenced by the radial profiles (RP) is shown in Column 5, and finally the total flux that we calculate from the AGN contribution (F$_{\mbox{\tiny AGN}}$) in Column 6.  The FWHM (Column 7) is determined using a Gaussian profile.

\begin{table*}
\centering
\caption{31.5 \um~photometric results from FORCAST.}
\label{Photometry}
\begin{tabular}{p{2cm}ccccccc}
\hline

Object		&	F$_{18''}$		&	PSF$_{\mbox{\tiny sub}}^{a}$		&	F$_{\mbox{\tiny PSF}}$	&	RP	&	F$_{\mbox{\tiny AGN}}$	&	FWHM	 		\\
			&	(Jy)			&	(\%)							&	(Jy)					&		(\%)		&	(Jy)		&	(arcsec)			\\					
\hline
MCG-5-23-16	&	2.43	$\pm$	0.20	&	75	&	1.64	$\pm$	0.25	&	97	&	1.59 $\pm$0.25		&	3.42	\\
Mrk 573		&	1.11 $\pm$       0.13	&	70	&	0.64	$\pm$	0.10	&	97	&	0.62	$\pm$0.10	&	3.81 \\
NGC 2110		&	1.25 $\pm$ 	0.15	&	80	&	0.86	$\pm$	0.12	&	94	&	0.81	$\pm$0.13	&	3.54	\\
NGC 2992	&	1.76	$\pm$	0.21	&	80	&	0.81	$\pm$	0.13	&	85	&	0.69	$\pm$0.11		&	3.38	\\
NGC 3081	&	0.96	$\pm$	0.12	&	85	&	0.80	$\pm$	0.14	&	90	&	0.72	$\pm$0.12	&	3.61	\\
NGC 3227	&	2.28	$\pm$	0.27	&	70	&	1.30	$\pm$	0.20	&	89	&	1.16	$\pm$0.19	&	3.97	\\
NGC 3281	&	2.68	$\pm$	0.32	&	70	&	2.50	$\pm$	0.33	&	100	&	2.50	$\pm$0.33	&	3.22	\\
NGC 4388	&	3.02	$\pm$	0.36	&	80	&	2.04	$\pm$	0.24	&	92	&	1.86	$\pm$0.30	&	3.68	\\
NGC 5506	&	4.11	$\pm$	0.49	&	85	&	3.66	$\pm$	0.50	&	96	&	3.53	$\pm$0.56	&	3.46	\\
NGC 7469	&	9.39 $\pm$	1.13	&	70	&	4.99 $\pm$	0.60	&	85	&	4.23	$\pm$0.68	&	4.19	\\
NGC 7674	&	1.75	$\pm$	0.18	&	70	&	1.62	$\pm$	0.24	&	100	&	1.62	$\pm$0.24	&	3.41	\\
\hline
\end{tabular}\\
$^{a}$Level of PSF subtraction. Column 2: The photometric flux within an 18 arcsec diameter aperture, Column 3,4: the per cent of PSF subtraction and the resulting flux, Column 5: percent flux from the radial profile (RP) analysis, Column 6: total flux after considering the PSF-scaling and radial profiles, Column 7: the FWHM of each AGN as determined using a Gaussian profile.
\end{table*}

\subsection{\textit{Spitzer}/IRS Spectral Decomposition}

To further quantify the flux contribution of AGN emission and star formation and/or diffuse extended dust emission within our PSF-scaling photometric measurements, we perform a spectral decomposition analysis.  The spectral decomposition also allows for a comparison of the PSF-scaling method at the resolution of SOFIA.  

Specifically, we use the routine DeblendIRS\footnote{The routine can be found at: \url{http://www.denebola.org/ahc/deblendIRS}} \citep{HC2015} which uses a linear combination of three spectral components to describe the total flux: 1) AGN emission (AGN), 2) star formation (PAH), 3) and host galaxy emission (STR).  A sample of templates for each component is provided by the software, then DeblendIRS tests all possible combinations.  Finally, it selects the combination that minimizes the $\chi^{2}$ and the coefficient of the root mean squared error (CRMSE) without the need to model extinction separately. 

The routine covers a wavelength range of $5 - 38$ \um, so we are able to effectively compare not only the PSF-scaling results, but also high spatial resolution $8 - 13$ \um~spectroscopy.  We obtained \textit{Spitzer} IRS spectra for the 11 AGN from Cornell Atlas of \textit{Spitzer}/IRS Sources (CASSIS; \citet{Lebouteiller}) with spectral coverage in the rest frame range $5 - 38$ \um; the spectrum for NGC 7674 has spectral coverage of $10 - 37$ \um.  Data were obtained in staring mode using the low resolution (R $\sim$ 60 -120) modules for most AGN.  Low resolution (R $\sim$ 600) data was not available for NGC 7674, thus high resolution was used.  For point-sources the optimal extraction is that which produces the best signal to noise ratio (SNR); this was the extraction used for 8 AGN.  For partially extended sources, a ``tapered column" extraction is used.  This was the preferred extraction for NGC 2992, NGC 3227, and NGC 4388.  

As previously stated, the decomposition separates the spectrum from \textit{Spitzer} into AGN, PAH, and stellar components using templates for each.  Many of the AGN templates are from high-redshift sources, so their spectra reach 37 \um~in the observed frame, but not in the rest frame.  For this reason, we discarded 90 of 189 templates in order to reach a spectral range beyond 31.5 \um.  Hence, for each AGN we used the spectral decomposition in the range $5 - 32$ \um~($10 - 32$ \um~for NGC 7674).  

\begin{figure*}
\includegraphics[scale=0.7]{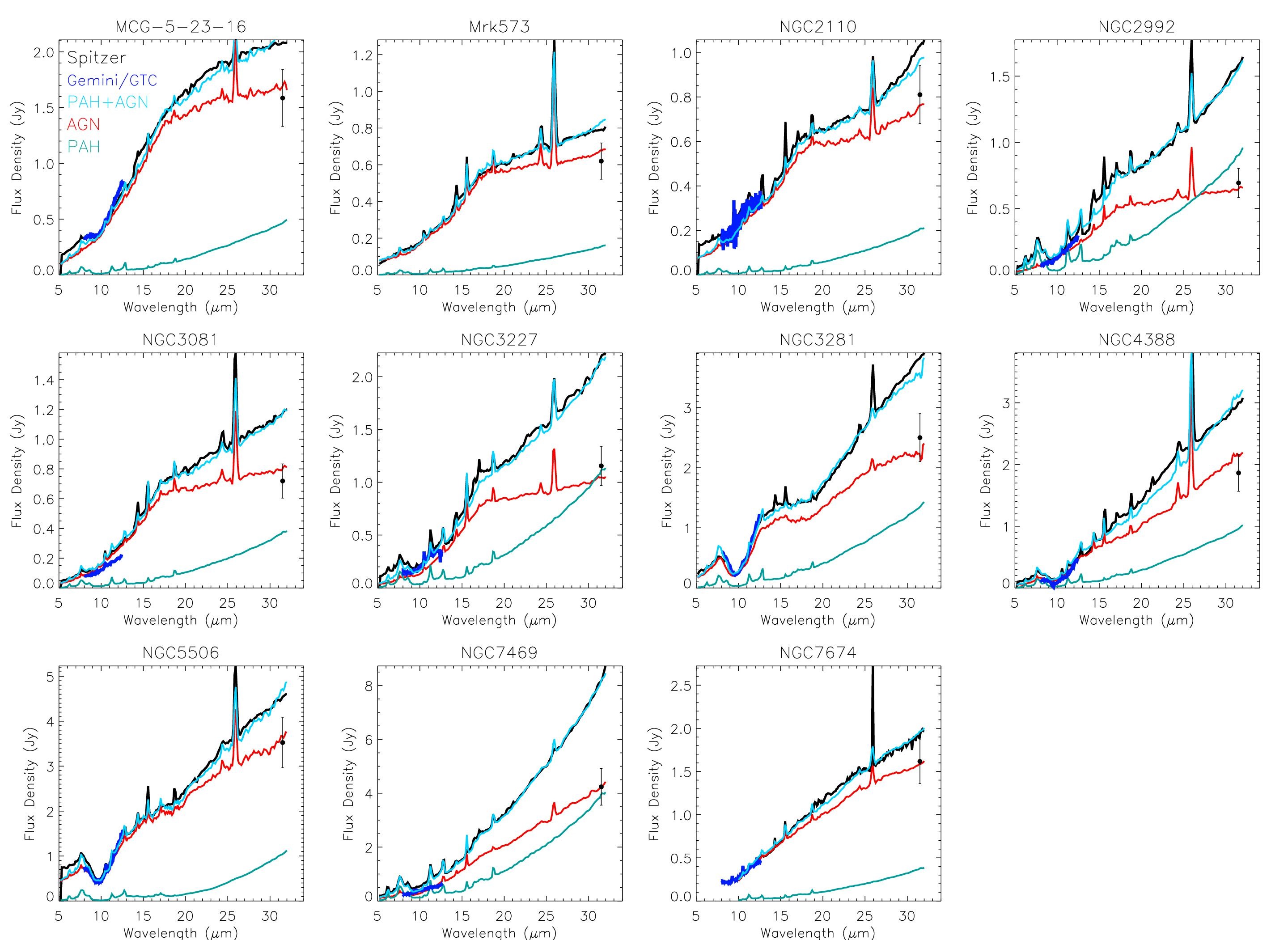}
\caption{Results of the DeblendIRS spectral decomposition of \textit{Spitzer}/IRS spectra plotted with the high spatial resolution spectroscopy (Table \ref{Spectroscopy}) and the SOFIA flux at 31.5 \um.  The solid black line is the Spitzer spectrum, while the green and red lines are the AGN and PAH contributions, respectively.  The cyan line indicates the addition of the AGN and PAH contributions.  The blue 8-13 \um~line is the ground-based high angular resolution spectroscopy, while the black dot is the SOFIA point with error bars.  We do not include the stellar component here because its contribution to the total flux is negligible ($<$0.1$\%$) }
\label{combined}
\end{figure*}

Figure \ref{combined} shows the $5 - 32$ \um~spectra of our AGN sample from CASSIS and the best fit to the spectrum using DeblendIRS.  The original spectrum from \textit{Spitzer}/CASSIS is shown in black, while the AGN and PAH components from DeblendIRS are shown in red and green, respectively.  The light blue (cyan) spectrum represents the addition of the AGN and PAH components, shown as a comparison to the original spectrum.  We do not show the stellar component because its contribution to the total flux is negligible ($<$0.1$\%$).  The subarcsecond resolution spectroscopy from Table \ref{Spectroscopy} is also shown for comparison in blue.  The AGN component from the spectral decomposition agrees with the results of the PSF-scaling, within the 1$\sigma$ uncertainty.  Thus we find the PSF-scaling photometric data to be an effective constraint to the nuclear SED modeling.

Table \ref{Contribution} shows a comparison between the AGN contribution at 31.5 \um~obtained from the PSF subtraction and from the spectral decomposition.  It can be seen from both methods that the objects in our sample are AGN dominated at 31.5 \um~with three exceptions: NGC 2992, NGC 3227, and NGC 7469.  All three were also shown to have some extension in their radial profiles (Figure \ref{rad_prof}).  

\begin{table*}
\centering
\caption{AGN contribution of PSF-scaling and spectral decomposition at 31.5 \um}
\label{Contribution}
\begin{tabular}{lcccccc}
\hline
		&	\multicolumn{3}{c}{(PSF-Scaling)}		&	\multicolumn{3}{c}{(Spectral Decomposition)} \\
Object	&	Total Flux		&	AGN Contr.	&	&	Total Flux	& 	AGN Contr.	\\
		&	 (Jy)		 	&		(Jy)			&	\%	&	(Jy)	&	(Jy)		&	 \%	\\
\hline
MCG-5-23-16	&	2.43$\pm$0.20		&	1.59$\pm$0.25		&	65	&	2.08$\pm$0.04	&	1.69$\pm$0.42	&	81	\\
Mrk 573		&	1.11$\pm$	0.13		&	0.62$\pm$0.10		&	56	&	0.79$\pm$0.01	&	0.68$\pm$0.17	&	86	\\
NGC 2110		&	1.25$\pm$0.15		&	0.81$\pm$0.13		&	65	&	1.03$\pm$0.03	&	0.76$\pm$0.19	&	74	\\
NGC 2992	&	1.76$\pm$0.21		&	0.69$\pm$0.11		&	39	&	1.58$\pm$0.03	&	0.65$\pm$0.16	&	41	\\
NGC 3081	&	0.96$\pm$0.12		&	0.72$\pm$0.12		&	75	&	1.17$\pm$0.02	&	0.80$\pm$0.20	&	68	\\
NGC	3227		&	2.28$\pm$0.27		&	1.16$\pm$0.19		&	51	&	2.19$\pm$0.03	&	1.05$\pm$0.26	&	48	\\
NGC 3281	&	2.68$\pm$0.32		&	2.50$\pm$0.33		&	93	&	3.82$\pm$0.05	&	2.23$\pm$0.70	&	58	\\
NGC 4388	&	3.02$\pm$0.36		&	1.86$\pm$0.30		&	62	&	3.00$\pm$0.04	&	2.14$\pm$0.54	&	71	\\
NGC 5506	&	4.11$\pm$	0.49		&	3.53$\pm$0.56		&	86	&	4.53$\pm$0.12	&	3.68$\pm$0.92	&	81	\\
NGC 7469	&	9.39$\pm$1.13		&	4.23$\pm$0.68		&	45	&	8.15$\pm$0.08	&	4.22$\pm$1.06	&	52	\\
NGC 7674	&	1.75$\pm$0.18		&	1.62$\pm$0.24		&	93	&	1.88$\pm$0.03	&	1.59$\pm$0.40	&	85	\\
\hline
\end{tabular} \\
Errors for the PSF-scaling are given in Table \ref{Photometry}.  There is a 25\% uncertainty in the fractional contribution of AGN to IRS flux \citep{HC2015}.

\end{table*}


\vspace{-0.7cm}

\section{NUCLEAR SED MODELING}
\label{MOD}

To investigate the effect of including the 31.5 \um~SOFIA/FORCAST photometric data on the torus parameters, we perform a nuclear SED fitting with and without the SOFIA data using the NIR and MIR photometry in Tables \ref{NIRphot} and \ref{MIRphot} and MIR spectroscopy in Table \ref{Spectroscopy}.  \citet{RA09,RA11,AH2011,Ichi2015} were successful in constraining torus model parameters for a large sample of AGN.  We aim to build on that work by using higher resolution data in the $30-40$ \um~wavelength range provided by SOFIA.  To infer physical properties of the torus using photometric and spectroscopic data, we use an interpolated version of the \textsc{Clumpy} torus models.  The Bayesian inference tool \textsc{BayesClumpy} \citep{AR2009} fits the IR SED using the uniform priors shown in the heading of Table \ref{clumpy}.

The \textsc{Clumpy} torus models are characterized by six parameters.  The model geometry consists of dust clouds with individual optical depths, $\tau$$_{v}$; an outer to inner torus radius ratio, $Y=R_{\mbox{\tiny out}}/R_{\mbox{\tiny in}}$; a radial distribution power law, $r^{-q}$; total number of clouds along an equatorial ray, \textit{N$_{0}$}; torus inclination angle, \textit{i}; and torus angular width, $\sigma$.  $R_{\mbox{\tiny in}}$ is set by the temperature of dust sublimation, $T_{\mbox{\tiny sub}}$$\sim$1500K and is computed using the AGN bolometric luminosity $R_{\mbox{\tiny in}}$ = $0.4 (L_{\mbox{\tiny bol}}/10^{45})^{0.5}$ pc.  For that reason, the modeled NIR to MIR SED is sensitive to the inner torus radius.  In addition to the six parameters, we estimate the torus sizes as $R_{\mbox{\tiny out}} = Y R_{\mbox{\tiny in}}$ pc, and the torus scale height, $H$, as $H = R_{\mbox{\tiny out}}\sin \sigma$ pc.    

The torus outer radius, $R_{\mbox{\tiny out}} = Y R_{\mbox{\tiny in}}$, is best constrained by adding FIR fluxes to the NIR/MIR SED \citep{AR2013,RA2014} since temperatures are generally cooler further away from the central engine.  The outer extent $Y$ is also highly coupled to the radial distribution.  In the case of steep radial distributions ($q$ = 2), $Y$ cannot be well constrained \citep{Nenkova2008b} due to the fact that most of the clouds are located close to the inner radius.  However, a flat distribution ($q$ = 0) provides a strong indication of $Y$ \citep{Thompson2009} since clouds are distributed uniformly throughout the torus volume.  \citet{RA2014} additionally show that NIR and MIR photometry is needed for $q$ to be constrained and MIR ($8-13$ \um) spectroscopy is needed in addition to NIR photometry to realistically constrain $Y$.

\begin{figure*}
\centering
\includegraphics[scale=0.75]{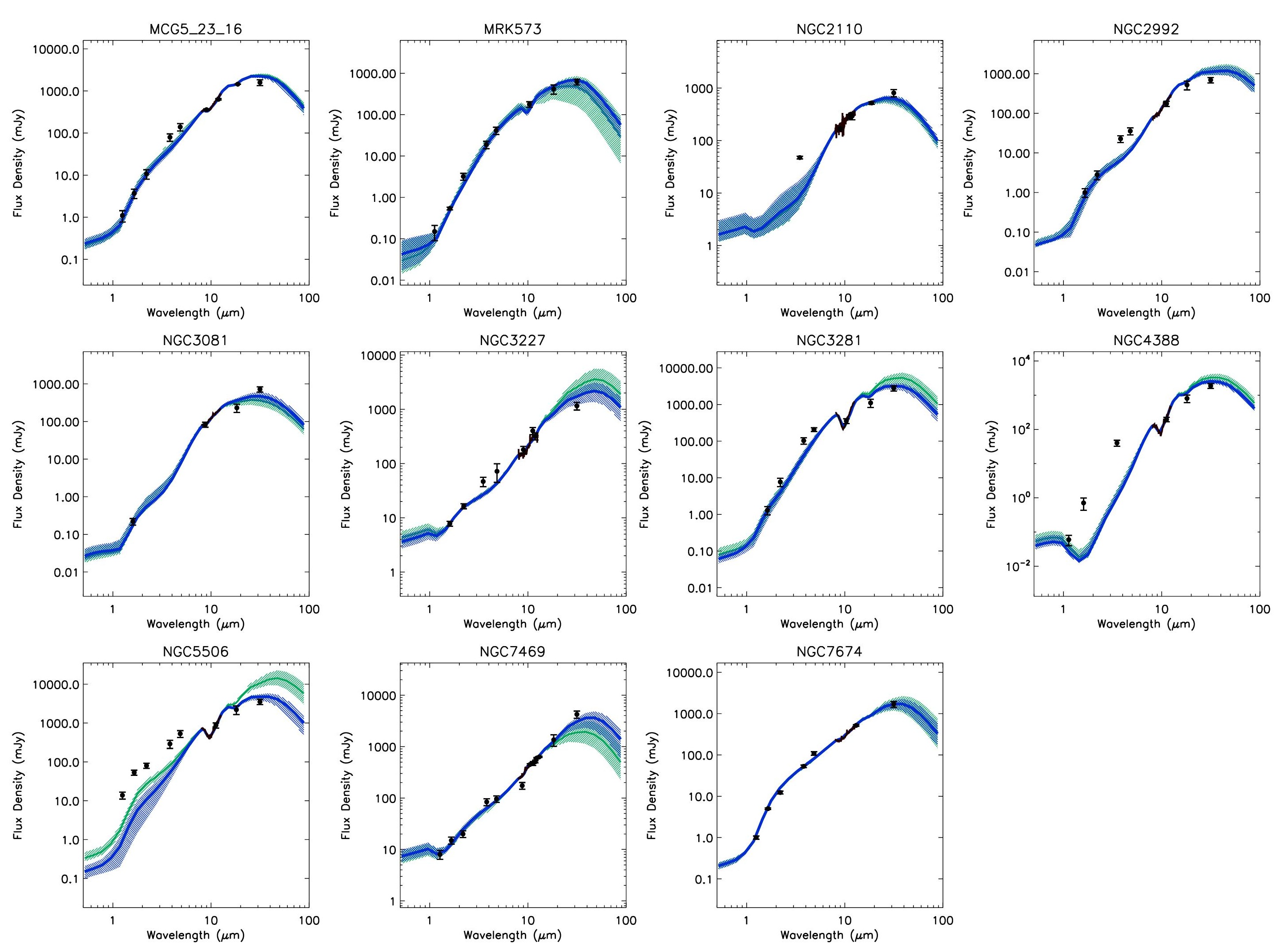}
\caption{\textsc{Clumpy} torus model fits. The SED photometric data (black solid filled dots), the models without (green solid lines) and with (blue solid line) the 31.5 \um~photometric data computed with the median value of the probability distribution of each parameter (Fig. \ref{posterior}), and the range of models compatible with a 68\% confidence interval (shaded area) are shown.}
\label{SED}
\end{figure*}

Figure \ref{SED} shows the models that best fit the SEDs, i.e. models described by the medians of the six posterior distributions resulting from the fit (Table \ref{clumpy}) without (green solid line) and with (blue solid line) the 31.5 \um~photometric data.  It can be seen from the figure that the data does not observationally show that the SED turnover occurs at wavelengths $\leq31.5$ \um.  The predicted wavelength of turnover emission lies between $30-50$ \um.  For most galaxies, the nuclear $1.2 - 31.5$ \um~emission is successfully fitted showing a reduction in the dispersion of clumpy torus models at wavelengths $>31.5$ \um~from those fits without including the 31.5 \um~fluxes.  These results support the assessment by \citet{AR2013} suggesting that photometric data from FORCAST provides significant constraining power for the \textsc{Clumpy} models. 

NGC 4388 and NGC 5506 both show a poor fitting to the $1 - 10$ \um~photometric data in the SED.  For both, the $8 - 13$ \um~spectroscopic data shows a strong 9.7 \um~silicate feature.  \citet{AH2011} show that, for NGC 5506, if the silicate feature were due to the torus only, it would be shallower and that the feature could be explained by some obscuration by an additional dust component.  This component could contaminate the NIR signal and cause the fitting discrepancy that is seen here.  NGC 4388 is a nearly edge-on Seyfert galaxy with a known host galaxy dust lane, which is reflected in its chaotic NIR spectrum \citep{Mason2015}.  The prominent dust lane could cause excess emission in the NIR seen here.  

Furthermore, NGC 4388 is known to have extended thermal emission due to a dusty ionization cone increasing in intensity from $\sim$10 to 18 \um~\citep{RA09}. This extended emission is also obvious in the SOFIA data extending farther out at $\sim$31 \um~indicating that the dusty narrow line region (NLR) may continue to have an increasing significance at longer wavelengths. Such dusty NLR's emitting at FIR wavelengths could be an additional source of contamination in the SOFIA data that requires further analysis.


\begin{table*}
\begin{minipage}{165mm}
\centering

\caption{Medians of Posterior Distributions}
\label{clumpy}
\begin{tabular}{lccccccccccccccc}
\hline

Object				&	$\sigma$					&	 $Y$								&	 $N_{\mbox{\tiny 0}}$		&	$q$							&	$\tau_{\mbox{\tiny V}}$			& $i$		&	$R_{\mbox{\tiny out}}$		&		$H$\\	 		
		&	[15\degree,75\degree]	&	[5,30]	&	[1,15]	&	[0,3]	&	[5,150]	&[0\degree,90\degree]		&	(pc)	&	(pc)	\\			
\hline
MCG-5-23-16	&	32$_{-5	}^{+7	}$	&	13$_{	-2	}^{	+2	}$	&	11$_{	-3	}^{	+2	}$	&	0.32$_{	-0.20	}^{	+0.39	}$	&	63$_{	-11	}^{	+13	}$	&	62$_{	-5	}^{	+4	}$	&	2.4$_{	-0.4	}^{	+0.4	}$	&	1.38$_{	-0.20	}^{	+0.26	}$	\\
	&	35$_{	-5	}^{	+7	}$	&	12$_{	-1	}^{	+2	}$	&11$_{	-3	}^{	+3	}$	&0.39	$_{	-0.25	}^{	+0.42	}$	&61	$_{	-11	}^{	+13	}$	&	61$_{	-5	}^{	+4	}$	&	2.2$_{	-0.2	}^{	+0.4	}$	&	1.38$_{	-0.18	}^{	+0.23	}$	\\
Mrk 573	&	50$_{	-18	}^{	+10	}$	&	17$_{	-6	}^{	+7	}$	&	8$_{	-3	}^{	+4	}$	&	1.12$_{	-0.65	}^{	+0.81	}$	&	24$_{	-6	}^{	+10	}$	&	51$_{	-33	}^{	+25	}$	&	4.2$_{	-1.6	}^{	+1.8	}$	&	3.40$_{	-1.05	}^{	+0.44	}$	\\
	&	46$_{	-16	}^{	+13	}$	&	19$_{	-6	}^{	+6	}$	&	7$_{	-3	}^{	+4	}$	&	0.61$_{	-0.38	}^{	+0.52	}$	&	27$_{	-7	}^{	+10	}$	&	63$_{	-37	}^{	+15	}$	&4.7	$_{	-1.6	}^{	+1.6	}$	&	3.57$_{	-1.09	}^{	+0.68	}$	\\
NGC 2110	&	57$_{	-9	}^{	+8	}$	&	19$_{	-6	}^{	+6	}$	&	9$_{	-2	}^{	+3	}$	&	2.71$_{	-0.26	}^{	+0.17	}$	&	140$_{	-12	}^{	+6	}$	&	44$_{	-24	}^{	+24	}$	&	1.9$_{	-0.6	}^{	+0.6	}$	&	1.70$_{	-0.19	}^{	+0.14	}$	\\
	&	55$_{	-9	}^{	+8	}$	&	21$_{	-6	}^{	+5	}$	&	9$_{	-3	}^{	+3	}$	&	2.60$_{	-0.30	}^{	+0.22	}$	&	141$_{	-10	}^{	+6	}$	&	40$_{	-24	}^{	+27	}$	&	2.1$_{	-0.6	}^{	+0.5	}$	&	1.83$_{	-0.22	}^{	+0.16	}$	\\
NGC 2992	&	52$_{	-7	}^{	+9	}$	&	11$_{	-1	}^{	+2	}$	&	13$_{	-2	}^{	+1	}$	&	0.17$_{	-0.12	}^{	+0.20	}$	& 133$_{	-10	}^{	+8	}$	&	39$_{	-12	}^{	+10	}$	&	1.0$_{	-0.1	}^{	+0.2	}$	&	0.87$_{	-0.07	}^{	+0.07	}$	\\
	&	52$_{	-7	}^{	+9	}$	&	10$_{	-1	}^{	+1	}$	&	13$_{	-2	}^{	+1	}$	&	0.18$_{	-0.12	}^{	+0.23	}$	&	130$_{	-16	}^{	+9	}$	&	41$_{	-15	}^{	+9	}$	&	0.9$_{	-0.1	}^{	+0.1	}$	&	0.78$_{	-0.08	}^{	+0.09	}$	\\
NGC 3081	&	62$_{	-6	}^{	+5	}$	&	10$_{	-3	}^{	+10	}$	&	12$_{	-2	}^{	+2	}$	&	2.65$_{	-0.42	}^{	+0.21	}$	&	126$_{	-19	}^{	+14	}$	&	63$_{	-21	}^{	+16	}$	&	0.4$_{	-0.1	}^{	+0.4	}$	&	0.35$_{	-0.02	}^{	+0.02	}$	\\
	&	62$_{	-7	}^{	+5	}$	&	20$_{	-7	}^{	+6	}$	&	12$_{	-2	}^{	+2	}$	&	2.65$_{	-0.25	}^{	+0.19	}$	&	125$_{	-18	}^{	+13	}$	&	58$_{	-25	}^{	+18	}$	&	0.8$_{	-0.3	}^{	+0.2	}$	&	0.71$_{	-0.05	}^{	+0.03	}$	\\
NGC 3227	&	55$_{	-4	}^{	+7	}$	&	21$_{	-2	}^{	+4	}$	&	12$_{	-2	}^{	+2	}$	&	0.07$_{	-0.04	}^{	+0.09	}$	&	148$_{	-3	}^{	+1	}$	&	23$_{	-5	}^{	+4	}$	&	1.1$_{	-0.1	}^{	+0.2	}$	&	0.97$_{	-0.05	}^{	+0.08	}$	\\
	&	53$_{	-4	}^{	+3	}$	&15	$_{	-1	}^{	+1	}$	&	13$_{	-2	}^{	+1	}$	&	0.05$_{	-0.02	}^{	+0.07	}$	&	148$_{	-3	}^{	+1	}$	&	13$_{	-8	}^{	+9	}$	&	0.8$_{	-0.1	}^{	+0.1	}$	&	0.68$_{	-0.04	}^{	+0.03	}$	\\
NGC 3281	&	65$_{	-5	}^{	+3	}$	&	20$_{	-3	}^{	+4	}$	&	12$_{	-3	}^{	+2	}$	&	0.22$_{	-0.15	}^{	+0.25	}$	&	38$_{	-5	}^{	+5	}$	&	25$_{	-10	}^{	+9	}$	&	2.1$_{	-0.3	}^{	+0.4	}$	&	1.97$_{	-0.09	}^{	+0.05	}$	\\
	&	66$_{	-3	}^{	+2	}$	&	15$_{	-2	}^{	+3	}$	&	13$_{	-2	}^{	+1	}$	&	0.19$_{	-0.13	}^{	+0.25	}$	&	35$_{	-5	}^{	+5	}$	&	14$_{	-8	}^{	+10	}$	&	1.5$_{	-0.2	}^{	+0.3	}$	&	1.49$_{	-0.04	}^{	+0.04	}$	\\
NGC 4388	&65	$_{	-5	}^{	+3	}$	&	22$_{	-3	}^{	+4	}$	&	14$_{	-1	}^{	+1	}$	&	1.16$_{	-0.25	}^{	+0.22	}$	&	41$_{	-5	}^{	+5	}$	&	79$_{	-9	}^{	+6	}$	&	1.2$_{	-0.2	}^{	+0.2	}$	&	1.15$_{	-0.05	}^{	+0.03	}$	\\
	&	66$_{	-4	}^{	+3	}$	&	18$_{	-2	}^{	+2	}$	&	14$_{	-1	}^{	+1	}$	&	1.02$_{	-0.26	}^{	+0.29	}$	&	36$_{	-4	}^{	+4	}$	&	82$_{	-7	}^{	+5	}$	&	1.0$_{	-0.1	}^{	+0.1	}$	&	0.95$_{	-0.03	}^{	+0.02	}$	\\
NGC 5506	&	60$_{	-3	}^{	+3	}$	&	23$_{	-2	}^{	+3	}$	&	13$_{	-2	}^{	+1	}$	&	0.07$_{	-0.04	}^{	+0.10	}$	&   82$_{	-5	}^{	+6	}$	&	16$_{	-5	}^{	+4	}$	&	3.5$_{	-0.3	}^{	+0.5	}$	&	3.17$_{	-0.10	}^{	+0.09	}$	\\
	&	60$_{	-5	}^{	+6	}$	&	13$_{	-2	}^{	+2	}$	&	11$_{	-2	}^{	+3	}$	&	0.22$_{	-0.15	}^{	+0.34	}$	&	57$_{	-8	}^{	+11	}$	&	25$_{	-8	}^{	+11	}$	&	1.9$_{	-0.3	}^{	+0.3	}$	&	1.79$_{	-0.10	}^{	+0.10	}$	\\
NGC 7469	&	33$_{	-5	}^{	+14	}$	&	16$_{	-4	}^{	+5	}$	&	6$_{	-2	}^{	+2	}$	&	0.32$_{	-0.21	}^{	+0.39	}$	&	135$_{	-13	}^{	+8	}$	&	58$_{	-9	}^{	+6	}$	&	3.6$_{	-1.0	}^{	+1.2	}$	&	2.10$_{	-0.29	}^{	+0.72	}$	\\
	&	34$_{	-5	}^{	+8	}$	&	26$_{	-4	}^{	+2	}$	&	9$_{	-3	}^{	+3	}$	&	0.19$_{	-0.12	}^{	+0.20	}$	&	133$_{	-13	}^{	+10	}$	&	50$_{	-6	}^{	+6	}$	&      6.0$_{	-1.0	}^{	+0.5	}$	&	3.50$_{	-0.47	}^{	+0.69	}$	\\
NGC 7674	&	26$_{	-5	}^{	+15	}$	&	22$_{	-3	}^{	+4	}$	&	9$_{	-4	}^{	+3	}$	&	0.21$_{	-0.13	}^{	+0.19	}$	&	68$_{	-18	}^{	+29	}$	&	56$_{	-9	}^{	+6	}$	&	8.4$_{	-1.2	}^{	+1.6	}$	&	3.86$_{	-0.70	}^{	+1.92	}$	\\
	&	25$_{	-5	}^{	+12	}$	&	22$_{	-3	}^{	+3	}$	&	9$_{	-4	}^{	+3	}$	&	0.20$_{	-0.13	}^{	+0.21	}$	&	66$_{	-17	}^{	+23	}$	&	56$_{	-8	}^{	+7	}$	&	8.4$_{	-1.2	}^{	+1.2	}$	&	3.72$_{	-0.71	}^{	+1.58	}$	\\
\hline

\end{tabular}\\
\end{minipage}
\textbf{Notes.}Medians of the posterior distributions of the \textsc{Clumpy} model parameters shown in Figure \ref{posterior} without (first row) and with (second row) the 31.5 \um~ photometric data. The input ranges considered for the fit are shown below the abbreviation of each parameter.
 \textsc{Clumpy} torus parameters: With of the angular distribution of clouds, $\sigma$; radial extent of the torus, $Y=R_{\mbox{\tiny in}}/R_{\mbox{\tiny out}}$; number of clouds along the radial equatorial direction, $N_{\mbox{\tiny 0}}$; power-law index of the radial density profile, $q$; optical depth per single cloud, $\tau_{\mbox{\tiny V}}$; inclination angle of the torus, $i$.
For the galaxies MCG-5-23-16 and NGC 5506 the input range of the $i$ parameter was assumed as a Gaussian prior into de computations, centered at 53\degree and 40\degree~respectively with a width of 10\degree~\citep{AH2011}.

\end{table*}



\begin{figure*}
\centering
\includegraphics[scale=.75]{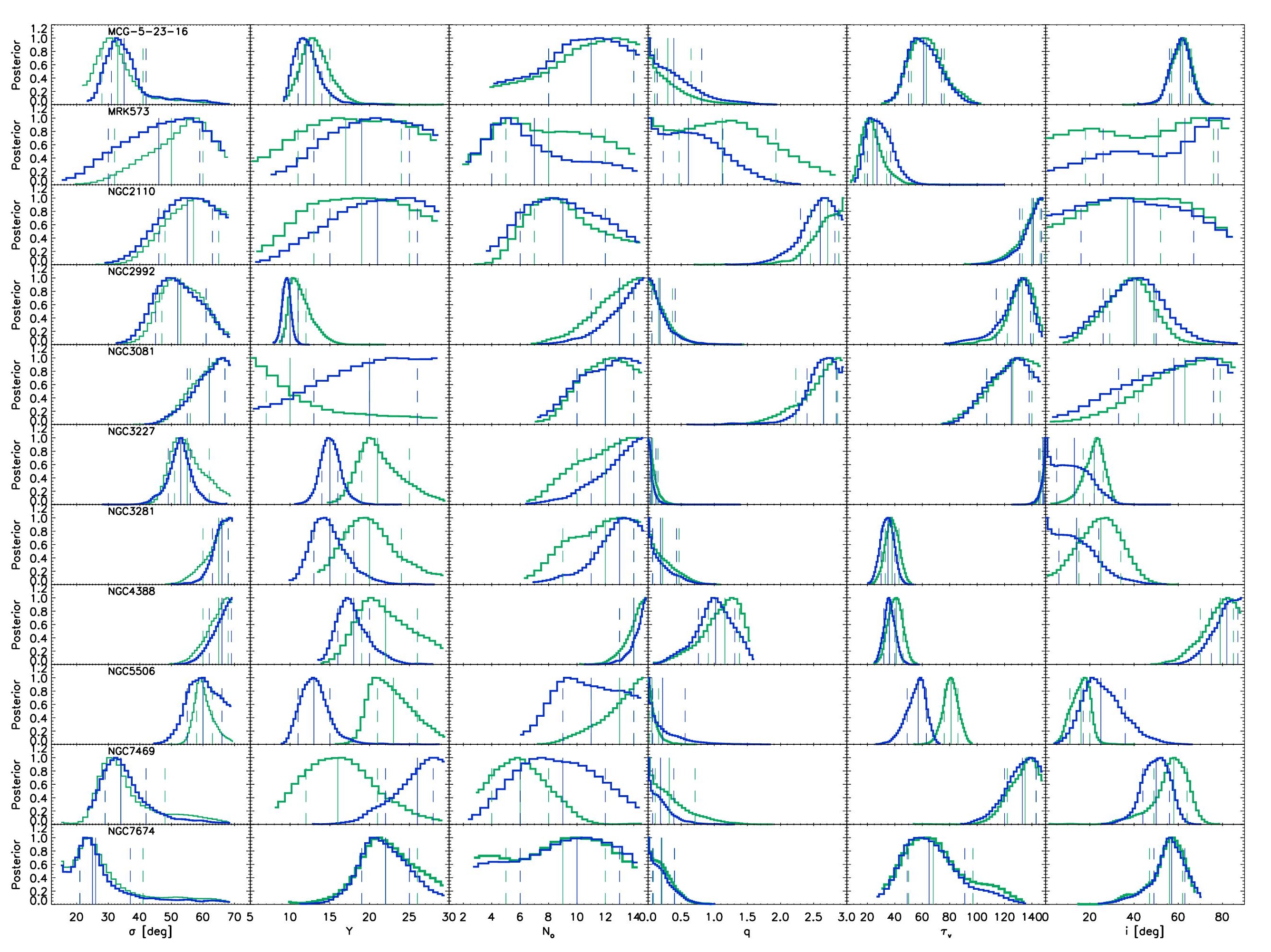}
\caption{Normalised posterior distributions with (blue thick solid line) and without (green thick solid line) the 31.5 \um~photometric data for each parameter derived from \textsc{Clumpy} torus models. Each column represents the marginal posteriors of the \textsc{Clumpy} parameters $\sigma$, $Y$, $N_{0}$, $q$, $\tau_{\mbox{\tiny V}}$, and $i$, respectively. The median (thin solid line) and $\pm$1$\sigma$ (dashed line) from Table \ref{clumpy} are shown.}
\label{posterior}

\end{figure*}



\vspace{-0.7cm}

\section{Discussion}
\label{DIS}

When dust emission at 31.5 \um~is taken into account in the SED, we find that the torus radial extent, $Y$, varies whereas the remainder of the parameters in the \textsc{BayesClumpy} output do not show any significant variation (Table \ref{clumpy}).  The radial extent shows a modification for 10 of the 11 objects; six (60\%) objects show a decrease in $Y$, whereas four (40\%) show an increase.  Of the 4 objects that show an increasing $Y$ value, one object does not have high resolution $8 - 13$ \um~spectroscopic data available (Mrk 573) and two significantly lack high-resolution NIR data (NGC 2110 and NGC 3081).  Hence, of the 8 AGN whose SED is well-sampled in NIR \textit{and} has spectroscopic data, 6 show a decreasing trend in the $Y$ parameter while one increases and one remains constant.  The relative change in $Y$ ranges from 0\% (NGC 7674) to 100\% (NGC 3081), with an average relative change of $\sim$30\% for the AGN in our sample.  We find torus outer radii in the range $<$1pc to 8.4 pc, which is consistent with the radii reported in the literature for Seyfert galaxies \citep{Jaffe2004,Packham2005,Tristram2007,Radomski2008}.  

The posterior distributions are shown in Figure \ref{posterior}.  It can be seen that the radial power law parameter, $q$, is generally less than 1.5.  In these cases, the outer to inner radius ratio $Y$ shows more constraint when including the 31.5 \um~data in the model.  It should be noted that, although Mrk 573 shows $q <$ 1.5 there is a large dispersion in $Y$ which we attribute to a lack of high angular resolution N-band spectroscopic data to include in the \textsc{BayesClumpy} fitting.  According to our results, the $q$ parameter generally peaks near zero, indicating that the clouds in the torus are most likely uniformly distributed throughout its volume and that the torus radial extent $Y$ can be constrained using the FIR nuclear fluxes presented here \citep{RA2014}.  For example, when 0 $\leq q \leq$ 0.5, 75\% to 85\% of clumps lie within 3/4 of the torus.  Since $q$ does not change significantly, but $Y$ does as shown in Table \ref{statistics}, we find that the same number of clouds are generally found in a smaller volume.  Hence, the physical interpretation is that clouds can be found more closely together, or the clouds have a higher mass density.   
       
For the two cases in which $q >$ 1.5 (NGC 2110 and NGC 3081), it is important to note that there is a significant lack of high spatial resolution NIR photometry for those AGN.  We note that a good sampling of NIR data is required in order to adequately describe $q$.  \citet{Nenkova2008b} point out that when $q$ = 2, the torus radial extent cannot be cannot be constrained, since the majority of the clouds are located relatively close to $R_{\mbox{\tiny in}}$ and therefore the SED is insensitive to $Y$.  This is one reason that for $q >$ 2 we see a large dispersion in $Y$ parameter values.  We further suggest that since $R_{\mbox{\tiny in}}$ is determined by the dust sublimation temperature, and since that temperature ($\sim$1500K) has a peak emission at $\sim$2 \um, high spatial resolution NIR data is needed to describe $Y$.  Hence, when $q >$ 2 \textit{and} there is a lack of NIR data, the torus radial extent $Y$ shows a large dispersion in values, in agreement with \citet{RA2014}.

\begin{figure*}
\centering
\includegraphics[scale=0.6]{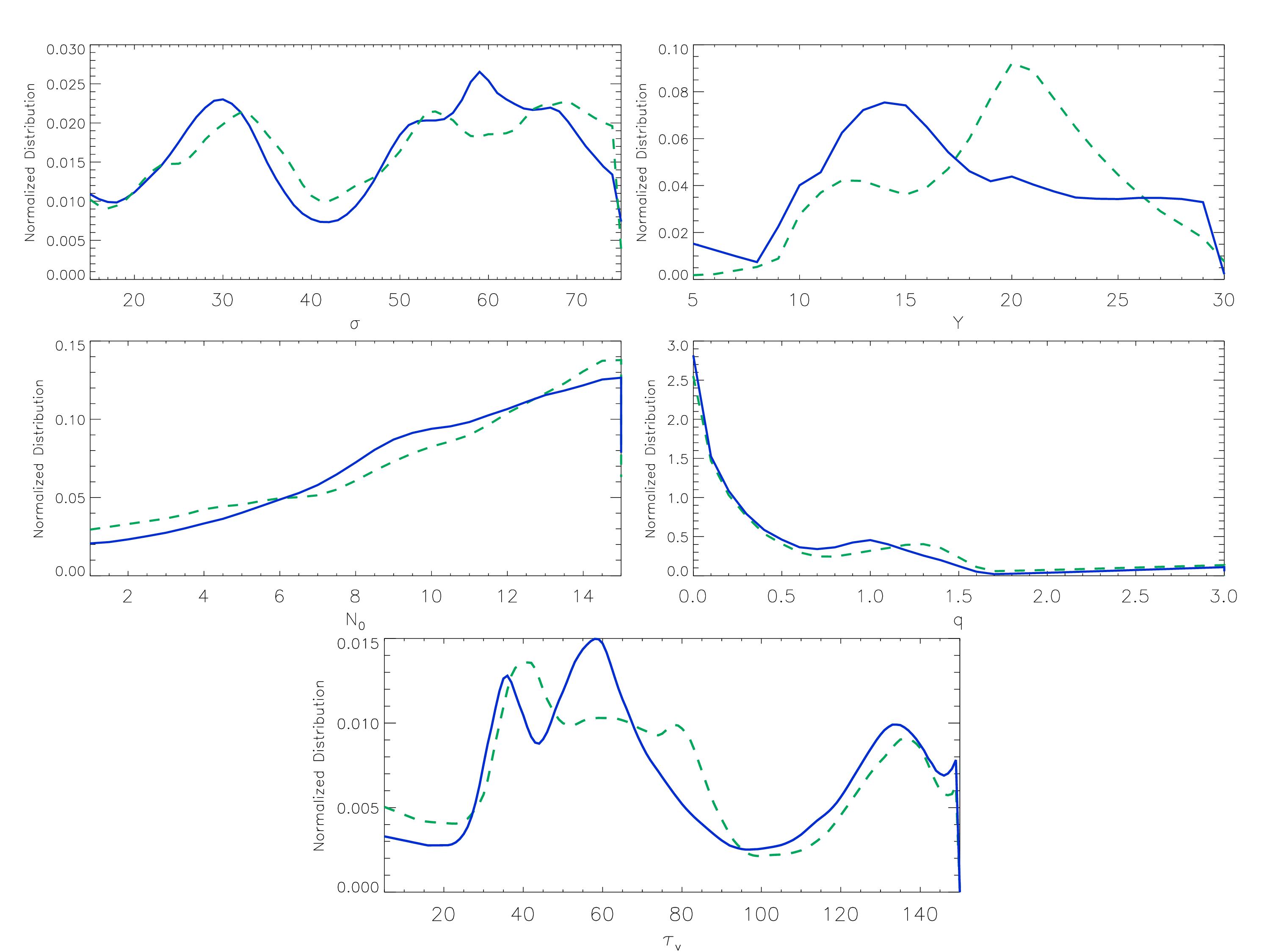}
\caption{Global posterior distibutions of the torus parameters without the SOFIA photometry (dashed green line) and with the SOFIA photometry (solid blue line).}
\label{norm_kld}
\end{figure*}

Figure \ref{norm_kld} shows the global posterior distributions of the physical parameters $\sigma$, $Y$, $N_{0}$, $q$, and $\tau_{\mbox{\tiny V}}$.  The inclination angle is not included because we used constrained prior distributions for MCG-5-23-16 and NGC 5506 as described in the notes of Table \ref{clumpy}.  Also, the global posteriors for Mrk 573, NGC 2110, or NGC 3081 were not used based on either their lack of spectroscopic or photometric NIR data.  The green dashed line represents the global posterior distribution not including the SOFIA data, while the solid blue line represents the global posterior including the SOFIA data.  The global distributions are essentially the same for $N_{0}$ and $q$, and very similar for $\sigma$ and $\tau_{\mbox{\tiny V}}$.  The most visually compelling difference is in the $Y$ parameter, where the distribution after adding the SOFIA data is lower than the distribution before adding the data.  

To quantitavely compare the differences in the global posterior distributions after the addition of the 31.5 \um~data, we calculate the Kullback-Leibler divergence \citep[KLD;][]{KLD}.  The KLD test compares the complete posterior distributions of the two sets of data rather than the median only.  For identical distributions, KLD = 0, while a large KLD value indicates that the posterior distributions are divergent.  Table \ref{statistics} gives the numerical KLD results for the remaining 8 AGN along with the median values calculated from the global posterior distributions (Figure \ref{norm_kld}) with and without the SOFIA photometry.  We find that when comparing our data, the KLD values tend to remain $<$0.1, with the highest value (for $Y$) being 0.306.  Given that the KLD value for $Y$ is at least an order of magnitude greater than those of the other parameters, $Y$ does present a statistically significant difference when comparing the parameters with and without the SOFIA data even though the global median values agree within their uncertainties.

\begin{table*}
\begin{minipage}{90mm}
\centering
\caption{Global posterior median values and Kullback-Liebler divergence results.}
\label{statistics}
\begin{tabular}{lccccc}
\hline
		&	$\sigma$	&	$Y$		&	N$_{0}$	&	$q$		&	$\tau_{\mbox{\tiny V}}$ \\
\hline
Median without SOFIA		&	51$_{-24	}^{+15}$	&	20$_{-7	}^{+4	}$	&	11$_{	-1	}^{+2	}$	&	0.4$_{-0.1	}^{+1.0	}$	& 	66$_{-30	}^{+63}$	\\
Median with SOFIA		&	50$_{-22	}^{+17}$	&	17$_{-5	}^{+7	}$	&	11$_{	-2	}^{+2	}$	&	0.4$_{-0.1	}^{+0.8	}$	&	66$_{-27}^{+66}$\\	
\hline
KLD results 	&	0.015	&	0.306	&	0.024	&	0.013	&	0.007	\\

\hline

\end{tabular}\\
\end{minipage}
\end{table*}

\defcitealias{Lira2013}{L13}

\citet{Stalevski2012} proposed a model in which the dust in the torus is distributed not as either homogeneous or clumpy, but as a two-phase medium in which high-density clumps are interspersed throughout a diffuse low-density medium.  Models like the two-phase torus models are already available in the literature, but further studies are needed to have a more physical interpretation of the torus.  \citet{Lira2013} (hereafter \citetalias{Lira2013}) also used the clumpy torus library of \citet{Nenkova2008,Nenkova2008b} to test the two-phase model on 27 Seyfert 2 objects.  The priors for $\sigma$, $N_{0}$, $q$, and $\tau_{\mbox{\tiny V}}$ in that sample are the same as in the current sample; however, the upper limit for $Y$ in \citetalias{Lira2013} is 100, where we restrict the upper limit to 30.

The general results agree quite well with our global posteriors.  \citetalias{Lira2013} find a large number of clouds $N_{0}$ $\gtrsim$ 10, where we find a global median $N_{0}$ = 11$_{	-2	}^{+2	}$, which is in excellent agreement.  For the opening angle, their general result is that $\sigma$ > 40\degree, which agrees with our global median, $\sigma$ = 50\degree$_{-22	}^{+17}$.  What is furthermore notable is a comparison between the two distributions of $\sigma$.  We show two peaks in our global posterior distribution, one $\sim$30\degree~and one $\sim$60\degree.  The model of \citetalias{Lira2013} also shows two peaks in its histogram, $\sim$15\degree and $\sim$60\degree.  While it is interesting to note that the three AGN in our sample with $\sigma$ < 40\degree have among the highest bolometric luminosities, we do not find a direct correlation between L$_{bol}$ and $\sigma$ throughout the whole sample.  A larger sample is needed in order to confirm that correlation.   \citetalias{Lira2013} also show a strong tendency for $q$ $\sim$ 0 and $Y$ $\lesssim$ 40, where we also show that $q$ tends to zero and $Y$=17$_{-5	}^{+7	}$.  Even though \citetalias{Lira2013} find higher values for the $Y$ parameter, they find typical torus radii between 0.1 - 5.0 pc, also in agreement with our modeling results. The only dissimilar parameter between the two data sets is $\tau_{\mbox{\tiny V}}$, which is smaller as determined by \citetalias{Lira2013} ($\lesssim$ 30) than in this data set (66$_{-27}^{+66}$).

\section{Conclusions}
\label{CON}

We present new 31.5 \um~imaging data from NASA's SOFIA/FORCAST for 11 nearby Seyfert galaxies.  To derive AGN-dominated fluxes within the 3.4 arcsec SOFIA aperture, we used the PSF scaling method of \citet{Radomski2003,RA09,Mason2012}.  We then performed a spectral decomposition using the routine of \citet{HC2015} to estimate the contribution of host galaxy components within SOFIA's FWHM and to verify the PSF-scaling results. We also compiled NIR and MIR fluxes from the literature based on previous works \citep{RA09, AH2011, Ichi2015}, and used the \textsc{Clumpy} torus models of \citet{Nenkova2008,Nenkova2008b} togther with a Bayesian approach \citep{AR2009} to fit the IR ($1.2-31.5$ \um) nuclear SEDs in order to constrain torus model parameters.   

The $1-31.5$ \um~SEDs presented here do not show a turnover below 31.5 \um.  The predicted turnover occurs between $30-50$ \um.  Further observations in the $32-40$ \um~regime would be beneficial in observationally finding the peak IR emission, thus providing further insight to the torus outer limit.  To further investigate this wavelength range, we have been allocated more observation time on SOFIA in 2016 to explore our current AGN sample, as well as 11 more objects, at a wavelength of 37.1 \um.  In the next decade, the 6.5-m Tokyo Atacama Observatory (TAO) may also present a unique opportunity to observe AGN in this wavelength range.  Its MIMIZUKU instrument will cover $2 - 38$ \um~with a spatial resolution of $1-2$ arcsec at 30 \um~\citep{Kamizuka2014}.

By including the 31.5 \um~photometric point in the SED, the model output for the radial extent is modified for 10 of the 11 objects.  Six galaxies (60\%) show a decrease in radial extent while four galaxies (40\%) show an increase.  The average value from the global posterior of $Y$ decreases from 20 to 17, with an average relative change of $\sim$30\% for the AGN in our sample, supporting the results of \citet{AR2013} which suggest that observations using FORCAST will provide substantial constraints to the \textsc{Clumpy} models.  We find torus outer radii in the range $<$1pc to 8.4 pc, consistent with interferometric and high angular resolution MIR observations \citep{Jaffe2004,Packham2005,Tristram2007,Radomski2008} and with recent sub-mm observations from ALMA \citep{GB2016}.

\vspace{-0.7cm}

\section*{Acknowledgments}

Based on observations made with the NASA/DLR Stratospheric Observatory for Infrared Astronomy (SOFIA). SOFIA is jointly operated by the Universities Space Research Association, Inc. (USRA), under NASA contract NAS2-97001, and the Deutsches SOFIA Institut (DSI) under DLR contract 50 OK 0901 to the University of Stuttgart. Financial support for this work was provided by NASA through award \#002\_35 and \#04\_0048 issued by USRA. E.L.R and C.P. acknowledge support from the University of Texas at San Antonio. C.P. acknowledges support from NSF-0904421 grant. C.R.A. is supported by a Ramón y Cajal Fellowship (RYC-2014-15779). AA-H acknowledges financial support from the Spanish Plan Nacional de Astronom\'ia y Astrofis\'ica under grants AYA2012-3144 which is partly funded by the FEDER program, and AYA2015-64346-C2-1-P. K.I. acknowledges support from JSPS Grant-in-Aid for Scientific Research (grant number 40756293).  N.A.L. is supported by the Gemini Observatory, which is operated by the Association of Universities for Research in Astronomy, Inc., on behalf of the international Gemini partnership of  Argentina, Australia, Brazil, Canada, Chile, and the United States of America.  IGB acknowledges financial support from the Instituto de Astrofísica de Canarias through Fundación La Caixa.  T.D-S. acknowledges support from ALMA-CONICYT project 31130005 and FONDECYT 1151239.  We would also like to acknowledge the contributions of Miguel Charcos-Llorens. 

\vspace{-0.8cm}










\bsp	
\label{lastpage}
\end{document}